\documentclass[aps,prd,nofootinbib]{revtex4}
\usepackage[colorlinks=true, pdfstartview=FitV, linkcolor=blue, citecolor=red, urlcolor=magenta]{hyperref}
\usepackage{graphicx}
\usepackage{latexsym}
\usepackage{amsmath}
\usepackage{amsfonts}
\usepackage{amssymb}
\usepackage{verbatim}
\begin{document}
\title{The Lorentz-violating effects in charged particle systems}

%%%%%%%%%%%%%%%%%%%%%%%%%%%%%%%%%%%%%%%%%%%%%%%%%%%%%%%%%%
\author{$^{1,2}$ E. Maciel}
\email{eugenio.maciel@df.ufcg.edu.br}

\author{$^{2}$ M. A. Anacleto}
\email{anacleto@df.ufcg.edu.br}

\author{$^{2}$ K.E.L.de Farias}
\email{klecio.limaf@gmail.com}

\author{$^{2,3}$ E. Passos}
\email{passos@df.ufcg.edu.br}

\affiliation{$^{1}$Unidade Acad\^emica de Engenharia de Produ\c{c}\~ ao, Universidade Federal de Campina Grande,\\
Caixa Postal 10071, 58540-000, Sum\'e, Para\'{\i}ba, Brazil.}

\affiliation{$^{2}$Unidade Acad\^emica de F\'{\i}sica, Universidade Federal de Campina Grande,\\
Caixa Postal 10071, 58429-900, Campina Grande, Para\'{\i}ba, Brazil.}

\affiliation{$^{3}$Unidade Acad\^emica de Matem\' atica, Universidade Federal de Campina Grande,\\
Caixa Postal 58429-970, Campina Grande, Para\'{\i}ba, Brazil.}

\begin{abstract}
We investigate the relativistic dynamics of a spin-half particle in the presence of a Lorentz-violating background within the framework of effective field theory. A modified Dirac Hamiltonian is considered, arising from a CPT-odd coupling involving the Lorentz-violating gauge tensor of the Standard Model Extension (SME). The velocity and effective force operators are derived from the Heisenberg equations of motion. Using Ehrenfest’s theorem and the correspondence principle, we obtain the classical limit of the dynamics and identify an effective force exhibiting a generalized Lorentz-force structure. This formalism is applied to a Penning trap system, known for its high-precision measurements of charged particle properties. Our analysis shows that the effective cyclotron frequency acquires a correction due to the Lorentz-violating term, leading to deviations in the particle trajectory and offering a potentially observable signature of Lorentz violation in precision experiments. By comparing our results with current bounds from high-precision Penning traps, we establish an upper limit on the Lorentz-violating coupling, 
 $\bar{g}k_{AF}\lesssim 2.32 \times 10^{4}\mathrm{eV}^{-1}$ corresponds to LIV effects. This bound is compatible with the interpretation of an effective Lorentz violation, consistent with current observational constraints, and it reinforces the phenomenological nature of the term under consideration, in agreement with previous analyses based on cosmological birefringence and photon propagation in a Lorentz-violating background.

%These results highlight the potential of relativistic effective models to probe new physics and reinforce the role of Penning traps as sensitive tools for testing Lorentz symmetry.
\end{abstract}
\pacs{11.15.-q, 11.10.Kk} 
\maketitle

%\vspace{1cm}
\pretolerance10000
\section{Introduction}
\label{Intro}
The search for a theory capable of unifying the fundamental interactions of nature seems to be one of the main objectives of current theoretical physicists. In this sense, theories beyond the Standard Model of elementary particle physics have shown themselves to be candidates for this ``\textit{new physics}" and this fact has motivated the investigation of scenarios in which some of the most fundamental principles of nature, among them, Lorentz symmetry, may be violated at high energy scales. In this context, many effective theories that incorporate Lorentz invariance violation (LIV) emerge as direct and phenomenologically consistent approaches to describe potential residual effects of a still unknown fundamental theory, such as quantum gravity or string theories \cite{mattingly2005modern,liberati2013tests,amelino2013quantum,kostelecky1989spontaneous}. Among the main effective theories, the Standard-Model Extension (SME), developed by Kostelecký and collaborators, provides a consistent, systematic and renormalizable framework that introduces LIV terms in the sectors of the Standard Model and gravity, in addition to, of course, relativistic quantum electrodynamics \cite{colladay1998lorentz,kostelecky2004gravity,kostelecky2022lorentz}. Numerous investigations across different physical systems seek to detect possible violations of this symmetry and other fundamental symmetries, ranging from highly energetic particles in accelerators to low-energy atomic systems, where such effects may be detectable thanks to advances in high-precision atomic spectroscopy techniques \cite{hohensee2013limits,bluhm2006overview,d2024irreducible,derevianko2018detecting,yerokhin2020self}.
% ${\color{blue}(k_{AF})_{\mu}}$
% {\color{blue}\bar{g}}
At the same time, the fermionic sector described from the Dirac theory plays a central role in describing the relativistic dynamics of spin-half particles, such as the electron, and is one of the most successful and applied theories in physics, widely used both in the description of fundamental theories and in approximations via effective theories, given its phenomenological potential. In this sense, a generalization via the introduction of effective terms that violate Lorentz invariance in the Lagrangian of the fermionic sector allows us to obtain the modified Dirac equation. Hence, it is possible to determine the effective Hamiltonian for the theory, which allows us to investigate modifications in physical observables that can be experimentally accessed with great precision \cite{altschul2005lorentz,Hohensee2013,diaz2012lorentz}. Following this way, it is possible to explore a modification of the Dirac equation with an additional effective term of the type $\bar{\psi} {(k_{AF})_\mu} \bar{F}^{\mu\nu} \gamma_\nu \psi$, where $\tilde{F}^{\mu\nu}$ is dual electromagnetic field-strength tensor and $ {(k_{AF})_\mu}$ 
is a constant four-vector. This background is responsible for the violation of Lorentz symmetry controlled by a dimensional constant ${\bar{g}}$, \cite{Belich:2004ng, Ribeiro:2007fe, Bakke:2011hg, casana2013new}. Based on this effective theory, we describe a complete analysis of the dynamics of a Dirac particle interacting with an external electromagnetic field. From the relativistic Hamiltonian, we used the Heisenberg formalism for the position and momentum operators, and determined explicit expressions for the velocity and force operators. The particle velocity operator remained unchanged in relation to the effective term, however, the force operator showed us an effective term controlled by the constant ${\bar{g}}$ and the vector ${{\bf k}}$. Armed with these results, it was possible to establish an analogy with classical physics between the obtained quantum effective force and the classical Lorentz-type d force. This analogy is justified by means of Ehrenfest's theorem, which relates the average evolution of quantum observables to the laws of classical motion, in agreement with the correspondence principle \cite{cohen1986quantum,merzbacher1998quantum,liboff2003introductory}.

To investigate the physical effects of this effective force, we propose its application to the context of Penning traps, highly precise experiments used to confine charged particles by means of static magnetic and electric fields. The proposal of the experiment was developed by F. M. Penning and later refined by Hans Dehmelt and has become one of the most important tools in high-precision experimental physics \cite{penning1936,dehmelt1990}. These devices have been known to allow ultrafine measurements of the charge-to-mass ratio, magnetic moments, and spectroscopic shifts of fundamental particles, such as electrons and antiprotons \cite{gabrielse1999precision,gabrielse2006new,sturm2014high,smorra2017parts,ulmer2015high,ulmer2018challenging}. We may implement simulations of the trajectory of particles subjected to the effective force in different regimes of Lorentz violation, revealing corrections in the cyclotron frequency that can, in principle, be detectable in precision experiments. Based on it, orders of magnitude can be estimated, and comparisons with recent experimental data allowed us to propose upper bounds for the LIV parameter. 

This paper is organized as follows. In section ~\ref{SC1}, we describe the modified electrodynamics, where it is possible to see the effective Lagrangian of the Chern-Simons type, showing the impacts of this theory on Maxwell's equations of classical electrodynamics. In section ~\ref{SC2}, we present the modified fermionic sector. In it, the Dirac equation will be explained, where we will obtain the effective Hamiltonian. Through it, the equations of motion for the position and velocity operators will be determined, and thus, we will have the velocity and force operators. In section ~\ref{SC3}, we will apply the results obtained in the Penning Trap experiment, where it will be possible to observe the impacts of the effective term on the cyclotronic frequency of the motion, altering the trajectory of the particle. Finally, in section ~\ref{SC4}, we present the final comments and perspectives.

%%%%%%%%%%%%%%%%%%%%%%%%%%%%%%%%%%%%%%%%%%%%%%%%%%%%%%%
\section{Electrodynamics with Lorentz Violation Background}
\label{SC1}
Namely, the modification of Maxwell equations involves adding to the four-dimensional Maxwell Lagrange density a Chern-Simons-like term \cite{Carroll:1989vb},

\begin{equation}\label{CFJ}
{\mathcal L}_{CS}= -\frac{1}{2} (k_{AF})_{\mu}\tilde{F}^{\mu\nu}A_{\nu}, 
\end{equation}
with $\tilde{F}^{\mu\nu} = \frac{1}{2}\varepsilon^{\mu\nu\alpha\beta}F_{\alpha\beta}$ being the dual electromagnetic field-strength tensor. Namely, the form of the CFJ term in 
Eq.(\ref{CFJ}) differs from the original definition, in which the four-vector, $(k_{AF})_{\mu}=(k_{0}, \bf{k})$, has mass dimension. In general, this original CFJ term naturally arises as part of the CPT-odd sector for SME, whose main objective is to describe all possible Lorentz and CPT violations in a systematic way in effective field theories (see also the Refs. \cite{jackiw1999radiatively,kostelecky2002signals,kostelecky2011data,colladay1997cpt}). The theory also provides the structure of the Standard Model in the low-energy limit. The presence of the vector $(k_{AF})_{\mu}$ selects a preferential direction in spacetime and violates the Lorentz invariance. In electromagnetic theory, it directly affects the propagation of electromagnetic waves and modifies Maxwell's equations in the following way:

\begin{equation}\label{eqsec011}
\partial_{\mu}F^{\mu\nu} = j^{\nu} + (k_{AF})_{\mu}\tilde{F}^{\mu\nu}.
\end{equation}
Note that the effects of the Chern-Simons-like term on the field equations is simply to replace of the source current four-vector, $j^{\nu}\to j^{\nu} +(k_{AF})_{\mu}\tilde{F}^{\mu\nu}$, such as the Maxwell equations in terms of components are
\begin{subequations}
\begin{equation}\label{eqsec012}
{\bf\nabla}\cdot{\bf E} = \rho - \big({\bf k}\cdot{\bf B}\big), 
\end{equation}
\begin{equation}\label{eqsec013}
{\bf\nabla}\times{\bf B} - \frac{\partial{\bf E}}{\partial t} = {\bf j}  - \big( k_{0}{\bf B} - {\bf k}\times{\bf E}\big).
\end{equation}
\end{subequations}
On the other hand, the homogeneous Maxwell equations given by $\partial_{\mu}\tilde{F}^{\mu\nu}=0$, remain unchanged:
\begin{subequations}
\begin{equation}\label{eqsec014}
{\bf\nabla}\cdot{\bf B} = 0, 
\end{equation}
\begin{equation}\label{eqsec015}
{\bf\nabla}\times{\bf E} + \frac{\partial{\bf B}}{\partial t}=0.
\end{equation}
\end{subequations}

From the set of Maxwell's equations above, we can conclude that the effective term provides modifications to Gauss's law for electromagnetism and Ampere-Maxwell's law, while Faraday's law and, of course, Gauss's law for magnetism remain unchanged. The scalar term ${\bf k}\cdot{\bf B}$ in Gauss's law acts as an effective electric charge induced from the magnetic field in conjunction with the preferred direction of ${\bf k}$. This fact is interesting since it implies that a magnetic field can directly contribute to the divergence of the electric field for certain reference frames. This result is incompatible with Maxwell's usual electrodynamics \cite{jackson2021classical,zangwill2013modern,schwinger2019classical}. In Ampere's law, the term $k_{0}{\bf B}$ has an analogy to a background current, which can be interpreted as an induced magnetically active medium. The term ${\bf k}\times{\bf E}$ has a structure similar to an anomalous Hall current. The presence of this term implies a direct coupling between the electric field and a preferential spatial direction. In general, the presence of these terms modifies the propagation of electromagnetic waves, in addition to the behavior of charges at rest or in motion, giving rise to effects such as the birefringence of light in a vacuum and unusual currents.
%%%%%%%%%%%%%%%%%%%%%%%%%%%%%%%%%%%%%%%%%%%%%%%%%%%%%
\section{Dirac Fermions in LIV Scenarios}
\label{SC2}
The dynamics of a relativistic particle interacting with an electromagnetic field is described based on the Dirac equation obtained from the Lagrangian that carries the interaction terms. We consider the term given by Eq.(\ref{CFJ}) as a non-minimal coupling involving fundamental Dirac fermions and the electromagnetic field in the LIV context. The four-dimensional Dirac Lagrangian density is given as
\begin{equation}
\label{DF1}
\mathcal{L}=\bar{\psi}\left(i \gamma^\mu \partial_\mu - m\right)\psi - q \bar{\psi} \gamma^\mu A_\mu \psi - \bar{g} \bar{\psi} (k_{AF})_\mu \tilde{F}^{\mu \nu} \gamma_\nu \psi.
\end{equation}
Here $\gamma^{\mu}$ are the Dirac matrices with $\gamma^{i}=\gamma^{0}\alpha^{i}$ and $\gamma^{0}=\beta$ where $i=1,2,3$ and $\bar{g}$  is the 
constant that measures the nonminimal coupling magnitude. Before proceeding, it is useful to clarify the mass dimension of the effective coupling appearing in Eq. (\ref{DF1}). Throughout this work we adopt natural units, $\hbar=c=1$, in which the Lagrangian density has mass dimension four. Since the Dirac field has dimension $[\psi]=3/2$, the dual electromagnetic tensor has dimension $[\widetilde F^{\mu\nu}]=2$, and the Lorentz-violating background coefficient satisfies $[(k_{AF})\mu]=1$, the operator $\bar{\psi}(k_{AF})_{\mu}\widetilde F^{\mu\nu}\gamma_{\nu}\psi$ has mass dimension $6$. Therefore, the coupling constant $(\bar g)$ must have mass dimension $-2$, implying that the effective combination $\bar g k_{AF}$ has dimension $-1$. Therefore, the phenomenological constraints obtained in this work are naturally expressed in units of $\mathrm{eV}^{-1}$.

The matrices $\alpha^{i}$ are defined in terms of the Pauli matrices $\sigma^{i}$ so that
\begin{equation}
\label{DF2}
\alpha^{i}=\left(\begin{array}{ll}
0 & \sigma^{i} \\
\sigma^{i} & 0
\end{array}\right).
\end{equation}
We emphasize that the effective term is CPT-even, however, it violates Lorentz symmetry and explicitly depends on the orientation of the four-vector $(k_{AF})_\mu$. The presence of this term leads to measurable physical effects, such as anisotropies in the propagation of fermions, shifts in energy levels in bound systems such as the hydrogen atom, and modifications in the equations of motion \cite{kostelecky1999constraints,Diaz2010}. Thus, the Dirac equation from (\ref{DF1}) is

\begin{equation}
\label{DF3}
\big(i\gamma^{\mu}\partial_{\mu} - q \gamma^{\mu}A_{\mu} -\bar{g}\, (k_{AF})_{\mu}\tilde{F}^{\mu\nu}\gamma_{\nu} - m\big)\psi = 0.
\end{equation}
It is important to emphasize that the Lorentz-violating background coefficient $(k_{AF})_\mu$ employed in the present work is inspired by the CFJ/SME framework, where it originally appears in the CPT-odd photon sector. However, the nonminimal fermionic interaction considered here, $\bar{\psi}(k_{AF})_\mu \widetilde{F}^{\mu\nu}\gamma_\nu \psi$, constitutes a distinct effective operator and should not be identified with the original CFJ term. Consequently, the discrete-symmetry properties of the fermionic interaction must be analyzed independently. Throughout this work, $(k_{AF})_\mu$ is interpreted as a fixed Lorentz-violating background entering an effective fermionic coupling,
rather than as the photon-sector CFJ operator itself. 

We use the following convention by introducing an antisymmetric rank tensor $2$, the field-strength tensor $F^{\mu\nu}$ and taking the components of the Dirac above, we will have the following structure:

\begin{equation}
\label{DF4}
i \frac{\partial \psi}{\partial t} = \big[{\bf \alpha}\cdot\big({\bf p} + {\bf \tilde{A}}) + m\beta  + \tilde{V} \big]\psi,
\end{equation}
where ${\bf \tilde{A}} =  q{\bf A} + \bar{g}\Big({\bf k}\times{\bf E}-k_{o}{\bf B}\Big)$ and $\tilde{V}= q\phi + \bar{g}\,{\bf k}\cdot{\bf B}$ are effective fields written in terms of the vector and scalar potentials of the usual electromagnetic dynamics ${\bf A}$ and $\phi$ respectively. Note that both LIV parameters $\bar{g}$ and ${\bf k}$ act on the effective fields. It should be highlighted that the electric and magnetic fields are defined in terms of the conventional potentials of electromagnetic theory according to the following relations: 

\begin{equation}
\label{DF5}
{\bf E}=-{\bf \nabla}\phi-\frac{\partial {\bf A}}{\partial t}\quad\mbox{}\quad
{\bf B}={\bf \nabla}\times{\bf A},
\end{equation}
with $q$ the charge of the particle. Hence, the modified Dirac Hamiltonian for a particle in an electromagnetic field with a LIV term may be obtained from Eqs. \eqref{DF4} and \eqref{DF5} together with the effective fields $\tilde{\bf{A}}$ and $\tilde{V}$, which leads us to the following expression
\begin{equation}\label{DF51}
\tilde{H}=\alpha\cdot({\bf p}+q{\bf A})+\beta m+q\phi+{\bar{g}}\Big[{\bf \alpha}\cdot\Big({{\bf k}}\times{\bf E}-{k_{0}}{\bf B}\Big) + {{\bf k}}\cdot{\bf B}\Big]
\end{equation}
\begin{equation}
\label{DF6}
    \tilde{H}=H_{D}+{\bar{g}}\Big[{\bf \alpha}\cdot\Big({{\bf k}}\times{\bf E}-{k_{0}}{\bf B}\Big) + {{\bf k}}\cdot{\bf B}\Big],
\end{equation}
where $H_{D}=\alpha\cdot({\bf p}+q{\bf A})+\beta m+q\phi$ is the usual Dirac hamiltonian for the particle in an electromagnetic field.
\subsection{Dynamics for a Particle in Lorentz Violation Background}

In this section, we will analyze the dynamics of a Dirac particle whose Hamiltonian is defined by equation (\ref{DF6}) and explore the contributions of the LIV term to the equations of motion for the operators. Investigating the dynamics of a particle is essentially analyzing its velocity and the force acting on it. However, in quantum mechanics, there is no concept of force, as in this context, a particle does not follow a trajectory. Therefore, when we claim to describe the dynamics of a particle from the point of view of relativistic quantum mechanics, we are referring to determining the equations for the velocity and force operators from the Heisenberg equations \cite{Sakurai1967,bjorken1964,greiner2008quantum}. Indeed, the correspondence principle together with Ehrenfest's theorem guarantees an analogy between the results of quantum mechanics and those of classical mechanics, thus allowing us to describe its dynamics.\footnote{It will be quite common throughout the text to treat quantities only as velocity and force, for example, but it is important to highlight that we are actually dealing with operators. For this reason, in order not to overload the notation, we are disregarding in our work the notation ``$(\ \hat{}\ )$" on quantities, quite common to describe operators.} In the Heisenberg picture, the velocity operator ${\bf v}$ is easily determined from the position operator ${\bf r}$ such that ${\bf v}=d{\bf r}/dt$ 
\begin{equation}
\label{S20}
    \frac{d{\bf r}}{dt}= \frac{1}{i}[{\bf r},{\tilde{H}}].
\end{equation}
The fact that the position operator ${\bf r}$ commutes with the fields ${\bf E}$, ${\bf B}$ and with the four-vector that controls the LIV, we can see that
\begin{equation}\label{S21}
  {\bf v} = \frac{1}{i}[{\bf r},{\tilde{H}}]={\bf \alpha}.
\end{equation}
The result of the equation above explicitly shows that the LIV term does not affect the velocity of the particle, giving the usual result \cite{Sakurai1967,bjorken1964,greiner2008quantum}. However, when we consider the modified kinetic momentum operator ${\bf \tilde{\Pi}}={\bf p}+ {\bf \tilde{A}}$, we find a different situation. Indeed, by applying the Heisenberg equation to the kinetic momentum operator, we find:
\begin{equation}\label{S22}
    \frac{d{\bf \tilde{\Pi}}}{dt}= \frac{1}{i}\Big[{\bf \tilde{\Pi}},\tilde{H}\Big]+\frac{\partial{\bf \tilde{\Pi}}}{\partial t}={\bf \tilde{F}},
\end{equation}
where ${\bf \tilde{F}}={\bf \tilde{E}}+{\bf v}\times{\bf \tilde{B}}$ is the modified force due to the effective term. 

The modified electric and magnetic fields are written in terms of the modified vector and scalar potentials ${\bf \tilde{E}}={\bf \nabla}\tilde{V}+\frac{\partial {\bf \tilde{A}}}{\partial t}$ and ${\bf \tilde{B}}={\nabla}\times{\bf \tilde{A}}$. Notice that this field is the same as in (\ref{DF5}). Consequently, explicitly have 
\begin{equation}\label{S23}
    {\bf \tilde{E}}=q{\bf E}+{\bar{g}\Big[{\bf \nabla}({\bf k}\cdot{\bf B})+\frac{\partial}{\partial t}({\bf k}\times{\bf E}-k_{0}{\bf B})\Big]},
\end{equation}
and
\begin{equation}\label{S24}
    {\bf \tilde{B}}=q{\bf B}+{\bar{g}\Big[{\bf \nabla}\times({{\bf k}}\times{\bf E})-{k_{0}}{\bf \nabla}\times{\bf B}\Big]}.
\end{equation}
The expressions (\ref{S23}) and (\ref{S24}) provide a curious analysis. Note that the modified electric field carries explicit magnetic field terms just as the magnetic field explicitly provides the electric field. This fact arises as a consequence of the coupling of these fields with the effective terms whose origin emerges from the modified Maxwell equations (\ref{eqsec012}) and (\ref{eqsec013}). Thus, from the modified fields (\ref{S23}) and (\ref{S24}) we can write the effective force (\ref{S22}) as
\begin{equation}\label{S25}
    {\bf \tilde{F}}={\bf F}+{\bar{g}\Bigg[\frac{k_{0}}{q}{\bf \nabla}\times{\bf F}+{\bf \nabla}({\bf k}\cdot{\bf B})+\frac{\partial}{\partial t}({\bf k}\times{\bf E})+{\bf v}\times\Big({\nabla}\times({\bf k}\times{\bf E})\Big)\Bigg]}.
\end{equation}
Here, the term ${\bf F}=q\left({\bf E}+{\bf v}\times{\bf B}\right)$ is the usual Lorentz force for electromagnetism. Considering the vector identities \cite{jackson2021classical}
\begin{equation}\label{S26}
  {{\bf \nabla}\left({\bf k}\cdot{\bf B}\right)=({\bf k}\cdot{\bf \nabla}){\bf B}+({\bf B}\cdot{\bf \nabla}){\bf k}+{\bf B}\times({\bf \nabla}\times{\bf k})+{\bf k}\times({\bf \nabla}\times{\bf B})},
\end{equation}
and 
\begin{equation}\label{S27}
    {{\bf v}\times\Big({\bf \nabla}\times({\bf k}\times{\bf E})\Big)={\bf v}\times\Big({\bf k}({\bf \nabla}\cdot{\bf E})-{\bf E}({\bf \nabla}\cdot{\bf k})+({\bf E}\cdot{\bf \nabla}){\bf k}-({\bf k}\cdot{\bf \nabla}){\bf E}\Big)}.
\end{equation}
Hence, by substituting these identities into Eq. (\ref{S25}) and considering the vector ${\bf k}$ as a constant, the effective force will therefore be 

\begin{eqnarray}\label{S28}
    {\bf \tilde{F}}&=&{\bf F}+\bar{g}\Bigg[\frac{k_{0}}{q}{\Big({\bf \nabla}\times{\bf F}}+{\bf k\cdot{\bf \nabla}\Big)}{\bf B}+{\bf k}\times{\bf j}+2{\bf k}\times\frac{\partial{\bf E}}{\partial t}+ \big({\bf k}\times({\bf k}\times{\bf E})\big)\\ \nonumber
    &+& {k_{0}({\bf k}\times{\bf B})+\rho{\bf v}\times{\bf k}- ({\bf v}\times{\bf k})({\bf k}\cdot{\bf B})-{\bf v}\times({\bf k}\cdot{\bf \nabla}){\bf E}\Bigg]}.
\end{eqnarray}

Note that if $\bar{g}=0$, we recover the usual dynamics of a charged particle in an electromagnetic field, as described by the Lorentz force \cite{greiner2008quantum,strange1998relativistic, dirac1928quantum}. In particular, the curl term of the Lorentz force that appears in the effective force above remains independent of the presence of the spatial vector ${\bf k}$. This term appears as a ``\textit{vorticity force}" to the usual Lorentz force, only disappearing when $\bar{g}=0$ or $k_{0}=0$. Unlike the usual Lorentz force, which depends locally on the fields ${\bf E}$ and ${\bf E}$, this term depends on the curl of the total force, which implies a sensitivity to the spatial non-uniformity of these fields. More details of this term will be discussed below. The other terms are corrections to electromagnetic dynamics due to Lorentz symmetry violation, which leads to non-standard effects such as modifications to the induction law or anisotropic couplings for the ${\bf E}$ and ${\bf B}$ fields. These effects become relevant at high-energy scales, where they may be measurable. 

 It is instructive to report the present Hamiltonian based approach to recent analyzes of relativistic spin dynamics in Lorentz-violating backgrounds. In particular, Ding, Kostelecký, and Vargas have recently investigated the relativistic spin precession of fermions in homogeneous background fields within a generalized Bargmann-Michel-Telegdi (BMT) framework \cite{ding2025relativistic}. In this work, the authors derives effective classical spin-precession equations directly from an extended BMT equation which is valid in arbitrary inertial frames. On the hand our approach proceeds from the underlying Dirac Hamiltonian and employs the Heisenberg equations of motion to obtain operator-valued expressions for the velocity, force, and spin dynamics. In the homogeneous and quasi-static field regime relevant to Penning-trap experiments, the resulting classical limit is structurally consistent with the generalized BMT description. A key distinction, however, is that the present formalism simultaneously yields modified orbital dynamics through an effective Lorentz force-like operator. This allows for a direct connection between Lorentz-violating background coefficients and experimentally accessible orbital observables, such as shifts in the effective cyclotron frequency. In this sense, the present analysis complements BMT-based treatments by providing a unified Hamiltonian framework for both spin and orbital effects in Penning traps.

%%%%%%%%%%%%%%%%%%%%%%%%%%%%%%%%%%%%%%%%%%%%%%%%%%%%%%%%%%%%%%%%%End Section
\section{Application: Penning trap Systems}
\label{SC3}
Penning trap experiments can be considered one of the most promising ways to test the viability of effective theories due to their high precision $10^{-11}$ and several practical applications \cite{bluhm1998cpt,kostelecky1999nonrelativistic}.
In general, a Penning trap is a device designed to trap charged particles, such as ions or electrons, in a region with an electric and magnetic field widely used in atomic and molecular physics, particle physics and metrology \cite{brown1986geonium,ghosh1995ion,hanneke2008new,abi2021measurement,logashenko2015measurement}. Important analyses in both experimental and theoretical fields, such as the determination of the charge-mass ratio $q/m$ and the determination of the factor $g$ for the electron, antimatter storage, and QED tests, make use of this experiment. The operating principle of this experiment consists of subjecting a charged particle to an intense static magnetic field ${\bf B}$ along the $z$ axis produced by a superconducting magnet or electromagnet and a quadrupolar electric field ${\bf E}$. The magnetic field restricts radial motion, while the electric field prevents axial runaway. The electric field is created by hyperbolic electrodes that generate an electric potential that confines the particles axially.

\begin{equation}\label{S291}
    \varphi(z,r)=\frac{V_{0}}{2d^{2}}\Big(z^{2}-\frac{r^{2}}{2}\Big),
\end{equation}
where $V_{0}$ is the applied potential and $d$ defines the geometry of trap with $r^{2}=x^{2}+y^{2}$.

The trajectory followed by a particle in a Penning Trap experiment is complex because it is composed of three different motions. The first is the cyclotron motion whose cyclotron frequency $\omega_{+}$ is modified by a second axial motion with axial frequency $\omega_{a}$. The last motion, defined as the magnetron motion, is characterized as a low-energy motion and describes a drift orbit defined by a characteristic frequency known as the magnetron frequency $\omega_{m}$. In general, due to the uniform magnetic field ${\bf{B}}=B_{z}\hat{k}$, a particle undergoes a circular motion in the $xy$-plane, similar to the usual motion of a charge in a uniform magnetic field with cyclotron frequency $\omega_{c}$. However, in the presence of the quadrupolar electric field $\bf{E}$, an axial perturbation arises at this frequency, so that the cyclotron frequency is
\begin{equation}\label{S281}
\omega_{+} = \frac{\omega_{c}}{2} + \sqrt{ \left( \frac{\omega_{c}}{2} \right)^2 - \frac{\omega_a^2}{2} }, \end{equation}
where $\omega_{c}=qB/m$ is the usual cyclotron frequency for a particle with velocity ${\bf v}$ in a magnetic field ${\bf B}$. For the axial motion, whose origin lies in the quadrupolar electric field {\bf E} that provides a restoring force along the $z$ axis, so that its motion is nearly harmonic along the $z$ axis. The frequency of this motion is therefore
\begin{equation}\label{S282}
\omega_a = \sqrt{ \frac{q V_0}{m d^2}}  . 
\end{equation}
Finally, magnetron motion occurs in the transverse plane $xy$, and has a slower rotation compared to cyclotron motion, characterized by a frequency $\omega_{m}$, defined by:
\begin{equation}\label{S283}
\omega_{m} = \frac{\omega_c}{2} - \sqrt{ \left( \frac{\omega_c}{2} \right)^2 - \frac{\omega_a^2}{2} }.
\end{equation}
Notice that this motion occurs in the opposite direction to cyclotron motion and is generally less energetic. Nevertheless, even with lower energy, it implies important experimental facts, mainly for the control and stability of trapped particles. This motion originates exclusively from the electric field ${\bf E}$. If $\omega_{a}=0$, it implies that $\omega_{m}=0$ in (\ref{S283}) and $\omega_{+}=\omega_{c}$ in (\ref{S281}), i.e., we recover the usual cyclotron motion. Note also that the condition
\begin{equation}
\omega_{c}=\omega_{+}+\omega_{m},
\end{equation}
is satisfied. It is important to note that the frequencies (\ref{S281}) to (\ref{S283}) are determined from the solutions of the differential equations that arise from the condition ${\bf E}=-\nabla\varphi$ with $\varphi$ defined by (\ref{S29}). For more details, see Ref. \cite{brown1986geonium}.
%%%%%%%%%%%%%%%%%%%%%%%%%%%%%%%%%%%%%%%%%%%%%%%%%%%%%%%%%%%%%%%%%%
\subsection{Penning Trap in LIV Scenario}
Namely, because $\omega_{c}$ is a much higher frequency than the axial $\omega_{a}$ and the magnetron $\omega_{m}$, any small change in $\omega_{c}$ caused by LIV has an amplified effect on differential or comparative measurements in a wide variety of experimental configurations. Thus, the cyclotron frequency provides the best response to the effective term, as well as the most directly affected and most precisely measurable of all the dynamical components of the system. This fact makes it the most powerful experimental channel for seeking evidence or imposing limits on the violation of Lorentz symmetry in confined charged particle systems. The present analysis assumes the idealized Penning-trap configuration with stationary quadrupolar fields. In more realistic experimental geometries, higher-order LIV corrections may induce weak couplings between radial and axial modes, potentially leading to small deviations from the invariant theorem. Such effects are expected to be subleading and remain as an interesting topic for future investigation. Therefore, although all frequencies can, in principle, be affected by LIV, the cyclotron frequency stands out as the most precise, direct, and reliable experimental route to investigate Lorentz symmetry violations in confined systems
\cite{abi2021measurement,logashenko2015measurement,ding2025relativistic}.

Now we will investigate the impacts caused by the LIV term in the Penning Trap experiment based on the effective force (\ref{S28}). We will perform our analysis in two different situations, namely for the four-vector $(k_{AF})^{\mu}$ for time-like and space-like vectors.
\begin{itemize}
    \item \textit{Time-like Vector ${{\bf k}=0}$ and ${k_{0}\neq0}$}
\end{itemize}

In this scenario, the special part of the vector that violates Lorentz invariance is zero, so the contribution to the effective force only includes the temporal part, from this point onward. In this sense, the only non-zero term of (\ref{S28}) is the first term inside the brackets, so that the force in this configuration is
\begin{equation}\label{S30}
    {\bf \tilde{F}}={\bf F}+{\bar{g}k_{0}\Bigg[\frac{1}{q}{\bf \nabla}\times{\bf F}\Bigg]}.
\end{equation}
Observing the equation above, we are tempted to consider significant effects on the dynamics of a particle in Penning traps, since this term carries an intriguing vorticity term arising from the usual Lorentz force coupled with the LIV term. In general, this term can be interpreted as a correction that depends on the spatial structure of the interaction, that is, as a kind of ``anomalous inductive force'' whose effect would be observable when the electromagnetic field allows non-negligible spatial variations. However, in Penning traps, the particles are confined by a constant magnetic field and a quadrupole electric field obtained from the potential (\ref{S291})
\begin{eqnarray}\label{S292}
    {\bf{E}}=\frac{V_{0}}{\text{2}d^{2}}\Big(x{\hat{i}}+y{\hat{j}}-2z{\hat{k}}\Big).
\end{eqnarray}

An additional comment is needed. Strictly speaking, the force appearing in Heisenberg's equation is an operator and therefore, the quantity $\nabla\times(\mathbf v\times\mathbf B)$ must be treated with caution before taking the classical limit. The effective force is fundamentally an operator relation, thus the correspondence with classical dynamics is only established after applying Ehrenfest's theorem and considering the evolution of expected values. In this semiclassical regime, appropriate for describing orbital motion in Penning traps, the expected value of the velocity behaves as a smooth dynamical variable, and the magnetic field can be considered uniform along the spatial extent of the particle's trajectory. Under these conditions, the expected value of the rotational contribution becomes:
\begin{equation}\label{S2921}
\left\langle\nabla\times(\mathbf v\times\mathbf B)
\right\rangle= (\mathbf B\cdot\nabla)\langle\mathbf v\rangle
-\mathbf B(\nabla\cdot\langle\mathbf v\rangle),
\end{equation}
where we use the fact that $(\nabla\mathbf B=0)$ for the ideal Penning trap configuration. Since cyclotron motion is described by a localized wave packet whose average velocity varies smoothly along the orbit and no spatially variable velocity field is generated in the trapping region, these contributions disappear in the dominant Ehrenfest approximation. Therefore, within the semiclassical description relevant to Penning trap experiments, the temporal contribution does not generate a dominant correction for the cyclotron dynamics. We emphasize, however, that this conclusion is based on the uniform field approximation and the use of expected values. A fully operational treatment in non-uniform electromagnetic configurations may yield additional contributions and deserves separate investigation.

Thus,  the curl of the Lorentz force is zero. In fact, take the curl of (\ref{S292}) and see that $\nabla\times\bf{E}=0$. In this case, we conclude that this term does not alter the dynamics of a particle in a Penning trap experiment. Terms of this nature are common in effective theories \cite{Belich2005,Ding2021,DingKostelecky2016}. Such terms are, in general, anisotropic and sensitive to the spatial structure of the field, which reinforces the fact that, in the absence of relevant spatial variations, as in the highly symmetric configurations of Penning traps, these effects are strongly suppressed or null.

\begin{itemize}
    \item \textit{Space-like Vector ${k_{0}=0}$ and ${{\bf k}\neq0}$}
\end{itemize}

Now we will do our analysis in the scenario that the vector is spacelike. Furthermore, the temporal part ${k_{0}=0}$ reduces the effective force (\ref{S28}) to
\begin{eqnarray}\label{S31}
    {\bf \tilde{F}}={\bf F}+{\bar{g}\Bigg[({\bf k}\cdot{\bf \nabla}){\bf B}+{\bf k}\times{\bf j}+2{\bf k}\times\frac{\partial{\bf E}}{\partial t}+\big({\bf k}\times({\bf k}\times{\bf E})\big)
    +\rho{\bf v}\times{\bf k}-({\bf v}\times{\bf k})({\bf k}\cdot{\bf B})-{\bf v}\times({\bf k}\cdot{\bf \nabla}){\bf E}\Bigg]}.
\end{eqnarray}
The force above shows the result of the effective force for the space-like case where ${k_{0}=0}$. Unlike the time-like case, there is a series of non-zero terms that reveal possible LIV effects originating from the space-like vector ${\bf{k}}$. However, for our system, this expression can be further simplified. In Penning traps, the quadrupolar electric field is considered stationary, that is, without temporal variations, so that $\partial\text{\bf{E}}/\partial t=0$. Furthermore, there are no net sources of charges or currents in the trap, that is, $\rho=\text{\bf{j}}=0$, because the trap operates in a vacuum with a single confined particle. Hence, taking these statements into account, the effective force can be rewritten as follows:
\begin{eqnarray}\label{S29}
    {\bf \tilde{F}}={\bf F}+{\bar{g}\Bigg[({\bf k}\cdot{\bf \nabla}){\bf B}+\bigg({\bf k}\times({\bf k}\times{\bf E})
    -({\bf v}\times{\bf k})({\bf k}\cdot{\bf B})\bigg)-{\bf v}\times({\bf k}\cdot{\bf \nabla}){\bf E}\Bigg]}.
\end{eqnarray}
{
Namely, our main analyses will be based on the above equation, and we highlight an important comment. The background that violates Lorentz symmetry is considered as fixed in an inertial frame centered on the Sun, according to the standard SME formalism (having no a priori geometric relationship with the magnetic field $\mathbf{B}$ defined in the laboratory \cite{Ding2021}). As will be shown below, there will be a correction for the cyclotron frequency, making it explicit that only the component of the background perpendicular to the axis of the magnetic field contributes, in the dominant order, to the transverse orbital dynamics associated with the cyclotron motion. We therefore restrict our analysis to the transverse projection $\mathbf{k}_\perp$ of the background relative to $\mathbf{B}$, which corresponds to the experimentally relevant and maximally sensitive configuration for the cyclotron frequency. In the most specific way, this occurs due to the fact that the components of the background in the laboratory are related to the components in the frame centered on the Sun by a time-dependent rotation, $k^{j}_{\rm lab}(t)=R^{jJ}(t)\,k^{J}_{\rm Sun}$, with $R^{jJ}(t)$ being the rotation matrix. In this framework, the transverse projection $k^{\rm lab}_\perp(t)$ naturally acquires a temporal dependence with constant terms and harmonic modulations at the Earth's sidereal frequency $\omega_\oplus$. For more details see Ref. \cite{Ding2021}.
} Thus, the focus of our analysis will be the cyclotron frequency $\omega_{c}$. 

Let us examine the effective force terms in (\ref{S29}) for typical Penning trap experiments. For this frequency, the relevant terms are those containing the velocity dependence $\bf{v}$, since the cyclotron motion results from the part of the equation of motion that contains the centripetal acceleration. A priori, we could disregard the terms that do not explicitly contain the velocity (the first and second terms in the brackets of (\ref{S29})), with this argument alone. However, some considerations are necessary regarding the non-use of these terms to determine the cyclotron frequency. First, in Penning traps, the magnetic field is constant, so the first term in brackets, which contains the derivative of the magnetic field, vanishes ${(\bf{k}\cdot\nabla)\bf{B}=0}$. The second term in brackets $\mathbf{k}\times(\mathbf{k}\times\mathbf{E})$ is quite interesting, although it does not contribute to the cyclotron motion. This term describes a conservative force since it does not depend on the velocity $\bf{v}$ and can eventually be incorporated into the effective potential. Because they do not induce rotational motion, they do not contribute to the centripetal force and, consequently, do not affect the cyclotron frequency. Thus, we excluded this term not because of its magnitude but because of its physical nature, since its contribution is independent of the particle velocity. In general, this term modifies the electrical confinement structure of the trap, and not the effective magnetic interaction responsible for the cyclotron motion, an analysis of this term is beyond the scope of this work. Terms like this are typical of effective theories \cite{Risse2007,MyersPospelov2003}.

Furthermore, the LIV terms involving electric field gradients and velocity vectors do not produce force components in the axial direction, leaving $\omega_a$ is unchanged. In fact, since the axial motion is purely oscillatory and does not involve vector components crossed with ${\bf{k}}$, the additional terms have no projection in that direction. For the magnetron frequency, it undergoes indirect corrections via modification of the cyclotron frequency $\omega_{c}$, since it depends on $\omega_{a}$ according to (\ref{S283}). Because $\omega_{m}$ has a very small absolute value, small changes in $\omega_c$ can be significantly reflected in $\omega_{m}$, although detectability is hampered by the lower resolution of the measurements in this range.
Taking these arguments into consideration, the effective force (\ref{S29}) can still be reduced to
\begin{eqnarray}\label{S30}
    {\bf \tilde{F}}=q{\bf E}+q\bf{v}\times\bf{\mathcal{B}},
\end{eqnarray}
where 
\begin{equation}\label{S31}
    {\bf{\mathcal{B}}}={\bf{B}}-{\frac{\bar{g}}{q}\Bigg[{\bf k}({\bf k}\cdot{\bf B})+({\bf k}\cdot{\bf \nabla}){\bf E}\Bigg]}.
\end{equation}
Since we are assuming the magic field is perpendicular to the background vector ${\bf{k}}$ so ${\bf{k}}$, so ${({\bf{k}}\cdot{\bf{B}})=0}$. Taking into account also that ${{\bf{k}}=k_{AF}\hat{i}}$, the second term in square brackets in (\ref{S31}) gives ${({\bf{k}}\cdot\nabla){\bf{E}}\sim \frac{V_{0}}{d^{2}}k_{AF}\hat{i}}$. With these considerations, the magnetic field assumes the following form
\begin{equation}
\mathcal{B}=B\hat{z}+\frac{\bar{g}k_{AF}V_{0}}{2qd^{2}}\hat{i}.
\end{equation}
This expression shows that the contribution induced by the Lorentz violation is perpendicular to the external magnetic field. Therefore, it does not alter the magnitude of ${\mathcal B}$ to the first order in $\bar{g}k_{AF}$, this magnitude must be expressed by
\begin{equation}
|\mathcal{B}|=\sqrt{B^{2}+\Big(\frac{\bar{g}k_{AF}V_{0}}{2qd^{2}}\Big)^{2}}.
\end{equation}
Taking an expansion up to first order, we obtain the expression
\begin{equation}
|\mathcal{B}|=B+\frac{1}{2B}\Big(\frac{\bar{g}k_{AF}V_{0}}{2qd^{2}}\Big)^{2}.
\end{equation}
In this sense, the effective cyclotron frequency $\tilde{\omega}_{c}=q|\mathcal{B}|/m$ is defined to be
\begin{equation}\label{S32}
    \tilde{\omega}_{c}\cong\omega_{c}+\frac{1}{2mBq}\Big(\frac{\bar{g}k_{AF}V_{0}}{2d^{2}}\Big)^{2}.
\end{equation}

We notice that the presence of the Lorentz-violating contribution modifies the equation of motion of a charged particle in a Penning trap. In particular, the cyclotron frequency undergoes a direct second-order correction as a function of the parameter ${\bar{g}k_{AF}}$, which controls the intensity of the Lorentz invariance violation. This correction stands out for its explicit dependence on the orientation and magnitude of ${\bf{k}}$. Thus the cyclotron frequency can vary depending on the orientation of the magnetic field in the laboratory, providing a clear signature of Lorentz breaking. 
Furthermore, by considering $\Delta \tilde{\omega}_{c} = \tilde{\omega}_{c} -\omega_{c}$, we can determine an upper limit to the Lorentz invariance violation effect as being
\begin{equation}
\label{S33}
\bar{g}k_{AF}\lesssim\frac{2d^{2}}{V_{0}}\sqrt{2mqB\Delta\tilde{\omega}_{c}}.
\end{equation}

A priori, we must be more rigorous with the expression (\ref{S33}) since, as we are considering the vector ${\bf{k}}$ fixed in a Sun centered reference frame, it explicitly shows a temporal dependence as mentioned earlier. However, from the practical point of view of this work, this dependence can be disregarded in the sensitivity estimation performed. In typical Penning trap experiments, the determination of the cyclotron frequency involves acquisition times significantly shorter than the sidereal period, or, alternatively, analysis procedures that perform temporal averaging on interventions in which harmonic variations are small. This fact allows the Lorentz violation effect to be planned, in a first approximation, by an average shift of the cyclotron frequency. 

Therefore, based on typical values for the Penning trap experiment, we can establish experimental constraints for the constant ${\bar{g}k}$. Moreover, considering quantities such as the electron mass $m=9.11 \times 10^{-31} \text{kg}\sim 5.11 \times 10^{5} \text{eV}$, the trap dimension $d=10^{-2}$, $\text{m}\sim 5.1 \times 10^{4}\ \text{eV}^{-1}$, $q \sim 0.303 $, the magnetic field $B=5\ \text{T}\sim976\ \text{eV}^{2}$ and the electric potential $V_{0}=10\ \text{eV}$ \cite{gabrielse2006new,gabrielse1999precision},  we arrive at the value
\begin{equation}\label{lim1}
\bar{g}k_{AF}\lesssim 2.32 \times 10^{4}\mathrm{eV}^{-1}.
\end{equation}
This value represents the upper limit for the magnitude of the of the parameter $\bar{g}k$ coupled with a charged particle in the context of an effective Hamiltonian obtained from the Dirac equation. Here we use the maximum measurement uncertainty $\Delta\tilde{\omega}_{c}=\tilde{\omega}_{c}-\omega_{c}=10^{-2}\ \text{rad}/\text{s} \sim 6.6 \times 10^{-18}\text{eV}$ for typical experiments of this trap \cite{gabrielse2006new,hanneke2008new}. Although this value is not competitive with the strongest existing astrophysical constraints on Lorentz-violating coefficients, this outcome is not unexpected. Astrophysical bounds are typically extracted from photon propagation over cosmological distances and benefit from enormous accumulated path lengths, whereas the present analysis is based on the dynamics of a single confined massive particle in a laboratory-scale system. Since the obtained bound explicitly depends on the particle mass, charge, and experimental frequency resolution, different confined systems such as antiprotons, heavy ions, or relativistic trapped particles may lead to distinct sensitivities to the effective LIV parameter $\bar g k_{AF}$.

Therefore, the significance of the present result should not be interpreted as providing the most stringent constraint on Lorentz violation. Instead, it demonstrates that Penning traps constitute a complementary laboratory platform capable of probing Lorentz-violating interactions through the orbital dynamics of confined charged particles \cite{colladay1997cpt,kostelecky1999constraints,Carroll:1989vb}. In particular, the present analysis reveals that the cyclotron frequency is sensitive to the spacetime orientation of the background vector and that Lorentz-violating effects manifest themselves through anisotropic modifications of the effective electromagnetic interaction. For these reasons, we interpret the present result as a conservative sensitivity estimate for a specific experimental channel rather than as a definitive bound on Lorentz-violating coefficients. The analysis nevertheless provides a consistent connection between the modified Dirac dynamics and experimentally accessible observables in Penning-trap systems.

%%%%%%%%%%%%%%%%%%%%%%%%%%%%%%%%%%%%%%%%%%%%%%%
\section{Final Remarks}
\label{SC4}
In this work, we investigate the dynamics of a spin-half particle in a uniform electromagnetic field with a background field defined by a constant quadrivector ${(k_{AF})_{\mu}}$ that violates Lorentz invariance, whose magnitude is controlled by a constant ${\bar{g}}$. We initially consider how Maxwell’s equations are modified by the CFJ term that violates Lorentz invariance. Then we determine the impacts of the effective term on the Dirac Lagrangian, where we expose the coupling with the electromagnetic field by means of the dual electromagnetic tensor. At this point, we obtain the modified Dirac equation, and from it, we find the effective Hamiltonian of the theory defined in terms of the effective vector and scalar powers. By means of the Heisenberg equations, it was possible to determine the equations of motion for the velocity and force operators and thus make an analogy with the classical results by means of Ehrenfest's theorem. We noticed that in the case of the velocity operator, the effective term did not bring any change, reproducing the velocity operator proposed by the usual Dirac theory. However, for the force operator, the effective term brought significant modifications in the particle dynamics, generalizing the usual Lorentz force with a term that depends explicitly on the effective term controlled by ${\bar{g}}k_{AF}$.

With these results at hand, we proceed to analyze by applied the theory to a Penning Trap type system of high precision and of great importance in atomic spectroscopy, in addition to being widely used to test effective theories beyond the standard model. Our proposal showed changes in the particle dynamics, mainly in the cyclotronic motion, where it was possible to estimate values of ${\bar{g}k_{AF}\leq10^{4}\text{eV}^{-1}}$. A careful analysis of the Penning-trap geometry shows that, for the configuration considered in this work, the Lorentz-violating correction to the effective magnetic field is perpendicular to the external magnetic field. Consequently, the leading correction to the cyclotron frequency appears only at second order in the effective parameter $\bar g k_{AF}$, yielding a quadratic frequency shift. This reinforce the importance of confined systems such as Penning traps as ideal platforms for testing extensions of the Standard Model with LIV, including the Standard-Model Extension (SME) \cite{kostelecky2002signals,kostelecky1999constraints} and theories with higher-dimensional operators, such as the Myers–Pospelov model \cite{MyersPospelov2003}. As a comparative analysis, although this limit is significantly less restrictive than the known astrophysical limits for minimum photon sector coefficients, such a difference is expected and physically consistent, since this is a laboratory test based on the dynamics of confined massive particles, sensitive only to specific background projections that violate Lorentz symmetry. In the context of Standard-Model Extension, limits of this nature are determined from the analysis of temporal modulations, in particular, lateral variations and their accuracy is directly linked to the spectral resolution and the time of acquisition of experimental data. Thus, the present result should be understood as a conservative estimate of sensitivity for this specific channel, which can be further improved in future analyses that explore long time series and more precise frequency measurement techniques, in complete analogy with other well-proposed tests of Lorentz symmetry violation in Penning traps.

The present model should be interpreted as a low-energy effective description of Lorentz-violating fermion dynamics. In this context, the physically relevant quantity is the effective composite parameter $\bar g k_{AF}$, and no specific ultraviolet completion is assumed. Possible radiative corrections induced by this higher-dimensional operator remain beyond the scope of the present phenomenological analysis. Thus, the parameter $\bar g k_{AF}$ considered in the present work should not be directly identified with the minimal SME photon-sector coefficient constrained by cosmological birefringence experiments. Instead, it represents an effective phenomenological coupling emerging from a non-minimal fermionic interaction. Furthermore, conservative anisotropic contributions associated with the term
$\mathbf{k}\times(\mathbf{k}\times\mathbf{E})$ remain present in the effective force and may modify the trapping potential, providing an additional channel through which Lorentz-violating effects could manifest themselves.

Finally, the results obtained demonstrate a significant modification in the dynamics of the confined particle, parameterized by the dimensionful effective coupling $\bar{g}k_{AF}$, which quantifies the intensity of the LIV. It is important to emphasize that this parameter should be interpreted as an effective phenomenological coupling in the fermion Hamiltonian, characterizing Lorentz-violating effects in the dynamics of confined particles, rather than as a fundamental minimal SME coefficient associated with the photon sector. This study, therefore, offers a bridge between effective models of new physics and high-precision experimental tests, reinforcing the role of Penning traps as promising experimental platforms to investigate signs of Lorentz-symmetry violation. As future perspectives, we propose investigating the influence of these terms on non-uniform field configurations and their potential signature in quantum interferometry and optical trap experiments.
%%%%%%%%%%%%%%%%%%%%%%%%%%%%%%%%%%%%%%%

{\acknowledgments} EM thanks to the Graduate Program in Physics at the Federal University of Campina Grande PPGF.  
MAA and EP acknowledge support from CNPq (Grant nos. 306398/2021-4, 304290/2020-3) and Paraiba State Research Foundation, Grant no. 0015/2019. KELF thanks the Paraiba State Research Foundation, FAPESQ.
% \bibliographystyle{ieeetr}
% \bibliography{references}

\end{document}